\newcommand{\vecA}{\mbox{\boldmath$A$}}
\newcommand{\vece}{\mbox{{\boldmath$\hat{r}$}}}
\newcommand{\vecex}{\mbox{{\boldmath$\hat{x}$}}}
\newcommand{\veceR}{\mbox{{\boldmath$\hat{R}$}}}
\newcommand{\vecnl}{\mbox{\boldmath$0$}}
\newcommand{\vecR}{\mbox{\boldmath$R$}}
\newcommand{\half}{\frac{1}{2}}
\newcommand{\vecr}{\mbox{\boldmath$r$}}
\newcommand{\vecv}{\mbox{\boldmath$v$}}
\newcommand{\veceta}{\mbox{\boldmath$\eta$}}
\newcommand{\vecbeta}{\mbox{\boldmath$\beta$}}
\newcommand{\lwp}{Li\'enard-Wiechert potentials}

\documentclass[aps,prl,twocolumn,groupedaddress,showpacs]{revtex4}

\begin{document}

%\preprint{}

\title{The exact Darwin Lagrangian}

% Please use the appropriate macro for each type of information

\author{Hanno Ess\'en}
\email{hanno@mech.kth.se} \homepage{
http://www.mech.kth.se/~hanno/ }
%\thanks{}
\affiliation{Department of Mechanics, KTH\\ SE-100 44 Stockholm,
Sweden}


\date{2007 July 25, corrections August}

\begin{abstract}
Darwin (1920) noted that when radiation can be neglected it should
be possible to eliminate the radiation degrees-of-freedom from the
action of classical electrodynamics and keep the discrete particle
degrees-of-freedom only. Darwin derived his  well known Lagrangian
by series expansion in $v/c$ keeping terms up to order $(v/c)^2$.
Since radiation is due to acceleration the assumption of low speed
should not be necessary. A Lagrangian is suggested that neglects
radiation without assuming low speed. It cures deficiencies of the
Darwin Lagrangian in the ultra-relativistic regime.
\end{abstract}

%\keywords{}

\maketitle


When radiation can be neglected the Lagrangian of classical
electrodynamics, putting $\vecbeta=\vecv/c$, can be written,
\begin{equation}\label{eq.L.ced}
L= \sum_a\left\{ - m_a c^2 \sqrt{ 1-\vecbeta_a^2 }+\frac{e_a}{2}
\left[\vecbeta_a \cdot \vecA(\vecr_a)- \phi(\vecr_a)
\right]\right\}.
\end{equation}
In 1920 Darwin \cite{darwin} expanded the \lwp\ to second order in
$\beta=v/c$ and thus found that,
\begin{equation}
 \phi (\vecr_a) = \sum_{b(\neq a)}
\frac{e_b}{|\vecr_a -\vecr_b|} =\sum_{b(\neq a)}
\frac{e_b}{r_{ba}},
\end{equation}
and (hats are used for unit vectors),
\begin{equation}
\label{eq.vec.pot.darw}
 \vecA (\vecr_a) = \sum_{b(\neq
a)} \frac{e_b [\vecbeta_b + (\vecbeta_b\cdot\vece_{ba})
\vece_{ba}] }{2 r_{ba}} .
\end{equation}
give the correct Lagrangian to this order. More recent derivations
can be found in a few textbooks
\cite{BKlandau2,BKanderson,BKjackson3}. In particular Jackson
\cite{BKjackson3} notes that using the Coulomb gauge
($\nabla\cdot\vecA=0$) makes the electrostatic Coulomb potential
$\phi$ exact and moves all approximation to the vector potential
$\vecA$ which obeys the inhomogeneous wave equation with the
transverse (divergence free) current as source. The Darwin
Lagrangian results when the term $c^{-2}\partial^2/\partial t^2$
in the d'Alembert operator is neglected so that the equation
becomes a Poisson equation.

The Darwin Lagrangian has been shown to be a unique
approximately relativistic Lagrangian (Woodcock and Havas
\cite{woodcock}, Kennedy \cite{kennedy}). It can be derived
from the Fokker-Wheeler-Feynman action-at-a-distance theory
(Anderson and Schiminovich \cite{anderson67}), and it is
useful in various fundamental studies of electrodynamics
\cite{coleman,crater2,essen05,boyer05}. The magnetic
interaction described by the Darwin Lagrangian is essential in
relativistic many-electron calculations as noted by Breit and
others \cite{breit29,sucher,deluca,yang&hirschfelder}. It has
found applications in nuclear physics
\cite{primakoff,balantekin}, and especially in plasma physics,
for numerical simulation
\cite{kaufman,nielson,ding,gibbons,leeww}, thermodynamics and
kinetics
\cite{krizan,essen96,essen&nordmark,alastuey,elboustani}, as
well fundamental theory \cite{mehra,essen99,essen06}. Barcons
and Lapiedra \cite{barcons} noted that the Darwin approach is
not valid for a relativistic plasma and therefore used a
different approach to its statistical mechanics.

Corrections to the Darwin Lagrangian have been discussed. Since a
system of particles with identical charge to mass ratio does not
dipole radiate a higher order expansion should be meaningful for
such systems \cite{golubenkov,dionysiou79,barker}. To that order,
however, acceleration inevitably enters and must be handled in
some way. Others have argued that since radiation is due to
acceleration, $v/c$ expansion is irrelevant, and further that
radiation can be negligible even if the particle speeds are
considerable (Trubnikov and Kosachev \cite{trubnikov4}, Frejlak
\cite{frejlak}). We will pursue that lead here.

One frequently encounters the statement that the Darwin approach
neglects retardation. This may be due to the fact that the,
nowadays best known, elegant derivation by Jackson
\cite{BKjackson3} hides the complications due to retardation.
Nevertheless it is wrong. The derivations by Darwin \cite{darwin}
and by Landau and Lifshitz \cite{BKlandau2} show that the
contribution of retardation to the Coulomb potential in the Lorenz
gauge, is quite large. The main acceleration dependent part,
however, vanishes either, as in Darwin's derivation, because it
gives a total time derivative term in the Lagrangian, or, as in
Landau and Lifshitz, because of a gauge transformation (to the
Coulomb gauge). Both these derivations also show that the velocity
dependent part of the retardation is handled exactly to order
$(v/c)^2$.

A natural idea that does {\em not} work is to assume constant
velocities and use the known exact \lwp\ for that case in
(\ref{eq.L.ced}). Darwin's original derivation shows that this
does not give the electric interaction to sufficiently
accuracy. It is important to note that gauge invariance (for a
review, see Jackson and Okun \cite{jackson&okun}), which is
valid for the exact theory, does not necessarily hold for
approximations. We therefore impose the Coulomb gauge (for a
recent discussion see Heras \cite{heras}) and then solve the
inhomogeneous wave equation for $\vecA$ assuming constant
velocities in the transverse current density. In this way one
treats the electric interaction exactly, neglects acceleration
in the solution for $\vecA$, but do not assume low speeds.

The constant velocity exact Coulomb gauge vector potential does
not seem to be well known. A special case was solved by Labarthe
\cite{labarthe}. The explicit general solution has recently been
published by Hnizdo \cite{hnizdo04p} who used a gauge
transformation function given by Jackson \cite{jackson} to find
it, starting from the corresponding \lwp. Denote by,
$\vecR=\vecr-\vecr'(t)$, the vector from the source particle at
$\vecr'(t)$, with charge $e$, to the field point $\vecr$, so that
$\vecbeta =\dot{\vecr}'(t)/c$. If we then put,
\begin{equation}\label{eq.eta.not.def}
\veceta = \veceR\times\vecbeta,
\end{equation}
Hnizdo's solution, which assumes the source particle to be at the
origin at time $t=0$ and to have constant velocity along the
$x$-axis, {\it i.e.} $\vecr'(t)=c\beta t\vecex$, can be written,
\begin{eqnarray}
  A_{Cx} &=& \beta\phi_L -(\phi_L - \phi_C)/\beta, \\
  A_{Cy} &=& \frac{y x}{y^2+z^2}(\phi_L - \phi_C)/\beta, \\
  A_{Cz} &=& \frac{z x}{y^2+z^2}(\phi_L - \phi_C)/\beta,
\end{eqnarray}
at $t=0$, so that $x, y, z$ are the components of $\vecR$. Here,
$\phi_C=e/R$, is the Coulomb potential, and, $\phi_L=\phi_C
/\sqrt{1-\eta^2}$, its Lorenz gauge form. One notes the identity
$1/(y^2+z^2)=(\beta/\eta)^2/R^2$. Using this, and that the only
relevant vectors are $\vecR$ and $\vecbeta$, one can, by
expressing everything in terms of these, or scalar and vector
products involving these, arrive at the coordinate independent
form,
\begin{equation}\label{eq.AC.constr5}
\vecA_C(\vecr) =\frac{e \left[ g(\eta^2) \vecbeta + h(\eta^2)
(\vecbeta\cdot\veceR) \veceR \right]}{R},
\end{equation}
of Hnizdo's solution. Here we have introduced the notation, and
the functions $g$ and $h$ are defined by,
\begin{equation}\label{eq.g.def.func}
g(x) \equiv \frac{1}{1+\sqrt{1-x}}\approx
\half+\frac{1}{8}x+\ldots,
\end{equation}
and,
\begin{equation}\label{eq.h.def.func}
h(x) \equiv  \frac{g(x) }{ \sqrt{1-x} }\approx
\half+\frac{3}{8}x+\ldots.
\end{equation}
Note that $g(1)=1$ but that $h$ diverges for $x=1$. From these
expansions it is clear that the leading terms give the vector
potentials (\ref{eq.vec.pot.darw}) of the Darwin Lagrangian.

The vector potential of the original Darwin Lagrangian is thus
recovered from (\ref{eq.AC.constr5}) when $\veceta=\vecnl$.
One notes that in the derivation of (\ref{eq.vec.pot.darw})
there was no need to assume that the velocity is constant
since the solution to the Poisson equation does not require
retardation, while it is necessary for solving the wave
equation. Hence the assumption of constant velocity. One
purpose of a Lagrangian is, after all, to find equations of
motion that determine the accelerations. If it is necessary to
know them beforehand a Lagrangian approach is pointless.

It is remarkable that an equivalent vector potential has been
found by Crater and Lusanna \cite{crater2} in a canonical
formalism. When the momenta (denoted $\vec{\kappa}$) of Eq.\
(5.28) of \cite{crater2} are replaced by $m \vecbeta/\sqrt{1 -
\beta^2}$ the expression (\ref{eq.AC.constr5}) is recovered.
The authors of \cite{crater2} use a relativistic phase space
formalism and assume that charges are anticommuting Grassmann
variables. In this way they treat the Pauli exclusion
principle semiclassically. Their Hamiltonian formalism, which
must entail the neglect of acceleration in an indirect way, is
mainly intended for treatment of bound states. The Hamiltonian
based on the ordinary Darwin Lagrangian (\ref{eq.L.ced}) is
discussed in \cite{essen&nordmark}.

The explicit expression for the interaction Lagrangian of two
particles that results when (\ref{eq.AC.constr5}) replaces
(\ref{eq.vec.pot.darw}) in (\ref{eq.L.ced}) is,
\begin{eqnarray}\nonumber
L_{12} = \frac{e_1 e_2}{r_{21}}\left[
\frac{g(\eta_1^2)+g(\eta_2^2)}{2} \vecbeta_1 \cdot\vecbeta_2 +\right. \\
\label{eq.L12.gen} \\ \nonumber \left.
\frac{h(\eta_1^2)+h(\eta_2^2)}{2} (\vecbeta_1 \cdot\vece_{21})
(\vecbeta_2 \cdot\vece_{21}) -1 \right],
\end{eqnarray}
where, $\eta_a^2 = (\vece_{ab}\times\vecbeta_a)^2$. We now
consider two special cases. If the velocity of a particle is
parallel to the inter-particle vector to another particle,
$\veceta=\vecnl$, so the Darwin interaction needs no correction in
these cases. Assuming that two particles have equal velocities
$\vecv_1 = \vecv_2 =\vecv$ parallel to $\vece_{21}$ we find the
the interaction term in (\ref{eq.L.ced}) gives,
\begin{equation}\label{eq.darw.interact.ahead}
L_{12}= \frac{e_1 e_2}{r_{21}}\frac{v^2}{c^2} - \frac{e_1 e_2 }{
r_{21}},
\end{equation}
and (\ref{eq.L12.gen}) gives the same result. One sees that
this term, and the corresponding force, goes to zero in the
ultra-relativistic limit $v \rightarrow c$. Now consider
instead the interaction of two particles that move with equal
velocities, $\vecv_1 = \vecv_2 =\vecv$, but side by side, so
that, $\vece_{21} \perp \vecv$. The interaction part of the
Lagrangian (\ref{eq.L.ced}) is then,
\begin{equation}\label{eq.darw.interact.side}
L_{12}=\half \frac{e_1 e_2}{r_{21}}\frac{v^2}{c^2} - \frac{e_1 e_2
}{ r_{21}}.
\end{equation}
One sees that even in the limit $v \rightarrow c$ the Coulomb
interaction dominates; the magnetic interaction can only
compensate for half of it. This is clearly wrong, however. It is
well known that in an ultra-relativistic beam the transverse
Lorentz force cancels the transverse Coulomb repulsion (see {\it
e.g.} \cite{BKwiedemann}).

Let us instead use the vector potential (\ref{eq.AC.constr5}) and
the corresponding interaction (\ref{eq.L12.gen}). We first note
that in this, side by side, case $\veceta^2 =v^2/c^2$ and thus
$\veceta^2=1$ in the limit $v \rightarrow c$. In this limit
$g(1)=1$ and $h$ diverges. The two scalar products in the second
term will, however, be zero and a simple investigation shows that
this compensates for the divergence of $h$, so that term does not
contribute. Finally we get,
\begin{equation}\label{eq.darw.interact.side.C}
L_{12}=g(v^2/c^2) \frac{e_1 e_2}{r_{21}}\frac{v^2}{c^2} -
\frac{e_1 e_2 }{ r_{21}},
\end{equation}
and in the limit $v \rightarrow c$ this term and the
corresponding force, are zero, as they should, when
(\ref{eq.L12.gen}) is used.

In conclusion, the Lagrangian obtained by using the exact
constant velocity Coulomb gauge vector potential
(\ref{eq.AC.constr5}), instead of the $\vecA$ used in
(\ref{eq.vec.pot.darw}), has been derived without assuming
that $v/c$ is small, only that accelerations are not needed in
estimating the Coulomb gauge vector potential. In this way all
velocity dependent retardation, and, as discussed above, also
the main part of the acceleration dependent retardation, is
accounted for. We have also shown that using this Lagrangian
we account correctly for the pinching of an ultra-relativistic
beam, something the original Darwin Lagrangian does not do.

% If in two-column mode, this environment will change to single-column
% format so that long equations can be displayed. Use
%\begin{widetext}
%\end{widetext}

% figures should be put into the text as floats.
% Use the graphics or graphicx packages (distributed with LaTeX2e)
% and the \includegraphics macro defined in those packages.
% See the LaTeX Graphics Companion by Michel Goosens, Sebastian Rahtz,

%\end{turnpage}

%\appendix
%\section{}

%\begin{acknowledgments}
%\end{acknowledgments}


%\bibliography{DarwinPapersRT,DarwinbooksRT,EMF_RT}

\end{document}